\documentstyle[12pt,aaspp4]{article}

\newcommand{\mg}{\mbox{ MG$\,0414$}}
\newcommand{\mga}{\mbox{ MG$\,0414+0534$}}

\newcommand{\etal}{\mbox{ et~al.}}

\newcommand\gtorder    
  {\mathrel{\raise.3ex\hbox{$>$}\mkern-14mu\lower0.6ex\hbox{$\sim$}}}
\newcommand\ltorder  
  {\mathrel{\raise.3ex\hbox{$<$}\mkern-14mu\lower0.6ex\hbox{$\sim$}}}

\slugcomment{To appear in \aj}

\lefthead{Falco et al.}
\righthead{HST Obs. and models of \mg}

\begin{document}

\title{HST Observations and Models of The Gravitational Lens 
System \mga\altaffilmark{1}}

\author{Emilio E. Falco, Joseph Leh\'ar, and 
Irwin I. Shapiro}
\affil{Harvard-Smithsonian Center for Astrophysics, 
60 Garden St., Cambridge, MA 02138}

\altaffiltext{1}{Based on observations made with the NASA/ESA Hubble
Space telescope, obtained at the Space Telescope Science Institute,
which is operated by AURA under NASA contract NAS5-26555}

\begin{abstract}
Quadruple gravitational lens systems offer the possibility of
measuring time delays for image pairs, 
microlensing effects, and extinction in
distant galaxies. Observations of these systems may 
be used to obtain estimates of H$_0$ and to study the various mass 
components of lens galaxies at high redshifts. 
With the Hubble Space Telescope, we have observed   
the reddest known gravitational lens system, \mga. 
We used WFPC2/PC1 to obtain 
deep, high-resolution images with two filters, F675W and F814W.
We present a detailed analysis 
of all of the components of \mg, as well as macrolens models. 
Our main results are:  
(1) confirmation that \mg\ is inescapably a gravitational lens system; 
(2) discovery of a blue arc connecting the 3 brightest 
images of the QSO central core;
(3) accurate positions and apparent brightnesses for
all 4 known images of the QSO central core and for the lens galaxy G; 
(4) a good representation
of the brightness distribution of G by 
elliptical isophotes with a De Vaucouleurs profile, characteristic 
of an elliptical galaxy;
(5) models that consist of simple elliptical potentials and account 
qualitatively, but not quantitatively, for the 
HST image positions, arc morphology and radio flux ratios for the 
images of the QSO central core; 
(6) a possible new test to distinguish between
reddening in the host galaxy of the QSO and in the lens galaxy,
based on future accurate measurements of spatial variations in
the color of the arc; and (7) the suggestion that 
microlensing is a plausible cause for  
the differences between the radio and optical flux ratios for
the brightest images, A1 and A2. 
Further observations and measurements such as of 
the redshift of the lens galaxy, 
can be used fruitfully to study microlensing for this system. 
\end{abstract}

\section{Introduction}

Among the most spectacular examples of gravitational lensing by isolated
galaxies are the quadruple systems of images, with characteristic 
sizes of a few arcseconds. 
\mga\ (hereafter \mg) was discovered by the MG radio survey
(see, e.g., Hewitt \etal\  1989), because of its 
distinctive ``quad'' radio morphology (Hewitt \etal\  1992, 
hereafter HTLSB92). 
The image configuration is the ``classic'' bright, close double (A1 and A2
separated by $0\farcs4$) 
assumed to straddle a critical curve, plus 
two additional, fainter images (B and C, each $\sim 2''$ from A1 and A2
and from each other). The redshift of each image is 
$z = 2.639 \pm 0.002$ (Lawrence \etal\  1995a, hereafter LEJT95; 
see also Katz \& Hewitt 1993, hereafter KH93; and Lawrence 1996).
The lens appears to be a single elliptical galaxy labeled G 
(Schechter \& Moore 1993, hereafter SM93). 
Its redshift remains undetermined;
attempts with the Keck I telescope 
initially appeared successful, but later were
shown to be a misidentification of 
Fe II lines at the redshift of the QSO (Lawrence \etal\ 1995b) as 
Na D lines at $z \sim 0.5$. 
The QSO images are 
exceedingly red, and extinction in the lens galaxy is
suspected of being significant (LEJT95). 

To further study \mg, we undertook a series of observations with the Hubble
Space Telescope (HST). 
We describe our observations in \S 2, as well as our reduction and
analysis of the ``raw'' HST images. In \S 3, we give our 
astrometry and photometry results for the components of the system.
We present a detailed analysis of the properties of the images, with 
models that account for most of their observed
properties in \S 4. Finally, our conclusions follow in \S 5.

\section{Observations}

\setcounter{footnote}{0}

We used WFPC2 to obtain 9 
deep, high-resolution images with two broad-band filters, F675W and F814W,
which are similar, respectively, to Johnson $R$ and $I$. 
We obtained these data 
on 8 November 1994, all in fine lock mode. We used
filter F675W for 4 of the exposures, one of 1800 sec and 3 of
2100 sec.
We used filter F814W for the other 5
equal-duration exposures of 2100 sec. 
%The exposures were obtained in successive orbits. 
The \mg\ system was located near
the center of PC1; unsaturated stellar objects on PC1 were aligned
from exposure to exposure with a resultant rms 
scatter of 0.13 pixels (each pixel is $\simeq 0\farcs0455$). 
We combined the exposures for each filter into a single image, to 
reject most of the numerous cosmic ray ``hits'' in our exposures.  
We used standard IRAF\footnote{IRAF (Image Reduction and Analysis Facility) is 
distributed by the National Optical Astronomy Observatories, which
are operated by the Association of Universities for Research in Astronomy,
Inc., under contract with the National Science Foundation.} 
tasks, which eliminated the majority of 
the ``hot'' pixels in the combined exposures. 
The combined images
are the means of 4 and 5 images for F675W and F814W, respectively, 
each scaled by its exposure time.
We will refer to these combined images as ``R'' and 
``I'' images, respectively. 

We detected the \mg\ system in each band, and found the 
components that were known from previous ground-based observations 
(HTLSB92; SM93; Angonin-Willaime et al. 1994, 
hereafter AVHM94) and from HST observations (Falco 1993).
Figures 1 and 
2 respectively show  $\sim 5\farcs8$ and 
$\sim 23\farcs3$ fields containing \mg\, extracted 
from the combined ``$R$'' images.  
The figures reveal the expected configuration of 4 point-like
images of the core, A1, A2, B and C, as well as the
lens galaxy G (Figure 1 labels these components).
In addition, our data reveal an arc that we interpret
(see \S 3) as lensed emission from the vicinity of the QSO. 
We detected a group of objects within
$\sim 1''$ of one another, $\sim 4''$ to the SW of the lens system. 
Each of the objects in this group is faint and extended
(see Figure 2). Hence, we assume it is a compact group of galaxies. 
Among the other objects in Figure 2 are 3 stars and other faint galaxies.

\section{Astrometry and Photometry}

The components of \mg\ are easily discerned in 
Figures 2 and 4. Figures 1 and 3 
%(Plate 1) 
show close-up views of the $R$ and $I$ images, respectively. 
The apparent ``fuzz'' surrounding 
the images of the core in the latter figure is
the low-level structure of the PC1 point-spread function (PSF). 
Although the images of the core are separated by only $\sim 2''$, 
the lensing galaxy was detected by Schechter \& Moore (SM93),  
under excellent observing conditions (FWHM=$0\farcs80$), 
and then confirmed by Falco et al. with WFPC1 (Falco 1993), 
and by Angonin-Willaime et al. (AVHM94).
As the figures show, we, too, detected the lens galaxy
with WFPC2. From our data, we are 
able to determine the spatial distribution of 
the lens galaxy's light emission at high 
significance levels. We also see clearly object X (SM93) 
near our detection limit, especially in $R$. 
Furthermore, one can easily see in both $R$ and $I$ 
(Figures 1 and 2) arc-like extended emission 
connecting A2 and B, and joining 
A1 and A2. The arc is unresolved in the direction from
the center of brightness of G. 
Such an arc can be explained as lensed emission from a
single, compact source near the QSO core, as we discuss in \S 4. 

We used IRAF to measure the pixel coordinates of the centers of brightness and 
the aperture magnitudes of objects with peak apparent magnitudes 
of 22.7 mag/arcsec$^2$ 
(24.2 mag/arcsec$^2$) above the sky level of the $I$ ($R$) image.
The absolute
coordinates, e.g., of B, agree with those of Katz \& Hewitt (KH93)
at the $0\farcs08$ level, consistent with their 
absolute positional uncertainty of $0\farcs1$. 
The images of the core and the galaxy were well above our 
detection thresholds above the sky level 
in each of $R$ and $I$. Object X was higher than the threshold only in 
$I$, although it is visible by eye in $R$.
We estimated its $R$ magnitude by fixing its position 
to that found in $I$, and comparing its brightness with that of C,
whose $R$ and $I$ magnitudes were well determined.

We converted the instrumental magnitudes to Johnson $R$ and
$I$ with the method of Holtzman \etal\ (1996).
We used their calibrated zero-points as appropriate for PC1 and for each
filter, and the corrections necessary for our nominal CCD gain. 
We set the magnitude scale for each band using color 
corrections derived from the ground-based colors of the 
images of the core and of G.
For the images of the core, the mean is $R-I \approx 1.8$; 
for G, $R-I \approx 1.6$, both based 
on AVHM94. 
Our magnitude estimates follow from the instrumental 
magnitudes within circular apertures 5 pixels in radius 
($\sim 0\farcs23$), except for that of G, which is clearly
not point-like (see below), and required a radius of 20 pixels. 
Tables 1 and 2 
show the coordinates and magnitudes of the images of the core, the
lens galaxy G, object X, the 3 brightest stars, and 
the faint, compact group of galaxies $\sim 4''$ SW of \mg. 
We used C as the origin of coordinates, because it is more
isolated than A1, A2 or B, and suffers less contamination from the arc.
The differences 
between the $R$ and $I$ coordinate offsets of the images of the core are 
$\leq 0\farcs015$. The maximum errors for centroiding the images of the 
core, as 
determined with IRAF, were 0\farcs02; we adopted these values as 
the standard errors for their coordinate
offsets in the tables. There is also very good agreement between our estimates
for the latter offsets, and the analogous estimates 
of KH93: the two sets differ by $\leq 0\farcs045$, the 
size of a PC1 pixel. 

Our magnitude estimates and those in AVHM94 for 
A1$+$A2 (unresolved in their observations), B, C and X differ
by $< 3\sigma$, using their errors, which are 
larger than ours by factors of $\sim 5-10$. Thus, our measurements should 
represent a substantial refinement of the estimates of the 
colors of the images of the core and of X that were
discussed in AVHM94. Our magnitude
estimates for G are fainter than those in AVHM94 by $\sim 0.7$ and 
$\sim 0.5$ mag in $R$ and $I$, respectively. These represent $\sim 5\sigma$
deviations, probably indicative of the (lower) 
accuracy that can be achieved by subtraction
of the images of the core in ground-based images of \mg. 
Table 3 shows the flux ratios A1/A2, (A1$+$A2)/B and C/B
for both of our pass bands, $R$ and $I$. Our ratios are consistent with
those of SM93, comparing their $I'$ band with our $I$, and 
also with those of AVHM94 in both $R$ and $I$. 

We estimated the apparent brightness of the longer part of the arc with a
polygonal aperture containing its emission, save for 
a PSF avoidance zone 10 pixels in radius from the centers
of brightness of the images of the 
core, which excluded any significant light in 
the PSF wings.
We could not obtain a useful 
estimate of the apparent brightness of the shorter portion
of the arc between A1 and A2, because the separation A1$-$A2 
is too small to yield magnitudes for the arc 
without severe contamination. 
We used IRAF to lay out a polygonal aperture along a $\sim 1''$ 
length of the arc
between images B and A2, thus minimizing contamination by A1, A2 and B.
We calculated $R$ and $I$ brightnesses with the same magnitude zero-points as 
for the images of the core, but avoided adding the color corrections, because 
we did not have ground-based estimates. We found 
$R_{\rm arc} = 24.3 \pm 0.3$, $I_{\rm arc} = 23.0 \pm 0.2$ and, hence, 
$R_{\rm arc}-I_{\rm arc} = 1.3 \pm 0.3$ 
mag and $R_{\rm core}-I_{\rm core} = 2.2 \pm 0.1$ mag; hence 
the arc is significantly (2$\sigma$ level) 
bluer than the QSO core.
We also subdivided our aperture into 4 equal, independent, 
adjacent rectangular 
apertures, to estimate color variation along the arc. 
We found no variation, within the 
standard error of 0.3 mag in our measurements.

We attempted to improve the resolution of our combined images by
deconvolution of the PSF in each band, 
to explore qualitative features of the images. 
We calculated PSFs appropriate to each 
band and to PC1 with Tiny Tim v. 4.0 (Krist 1996). 
Figure 5 shows the results of MEM deconvolution with these PSFs. 
The deconvolved brightness levels are shown as a contour plot
for the $R$ image, where the arc stands out most clearly. 
This figure shows that the images of the core 
were sharpened to point-like sources. The longer segment of
the arc joining B and A1 remains faint, but 
the shorter segment joining A1 and A2 can now be seen very clearly.
A very small, faint extension also 
appears to emanate from image C, toward its North,
similar to that seen in EVN VLBI observations (Patnaik \& Porcas 1996; 
henceforth PP96). 

We fitted 
elliptical isophotes to the brightness  
distribution of G in $I$, where it is most
clearly defined in our combined images. We adjusted the 
pixel coordinates of the centers of the isophotes, as well as 
their ellipticities $e_i = 1 - b_i/a_i$, where $a_i$ and $b_i$ are
the major and minor semi-axes of the $i^{th}$ isophote, respectively, 
and the position angles of the 
$a_i$. The coordinate offsets of the center of brightness of 
G from C are listed in Tables 1 and 2. 
The standard errors in these quantities are $\sim 0\farcs05$, 
as estimated from the isophote fitting. 
The mean ellipticity of the fitted isophotes is 
$e_G = 0.20 \pm 0.02$, and their mean position angle (E of N) on the sky 
is $\Phi_G = 71^\circ \pm 5^\circ$. The main contributor to the standard 
errors for these elliptical isophotes is the 
faintness of the outer regions of G.
The residuals from the subtraction of
the elliptical isophote model from the image
were consistent with sky noise, as expected for a good fit.

We also fitted profiles to the distribution 
of surface brightness of G as a function
of the mean semi-major axis of its elliptical isophotes. 
We convolved the model profiles with the azimuthally-averaged 
profile of the PSF that we created with Tiny Tim 
v. 4.0, as above, for the $I$ data.
Because we used only an azimuthal average of the PSF, the procedure
is not significantly sensitive to the details of the PSF wings.
We found a good fit ($\chi^2/DOF \sim 1.7$, where $DOF=38$ is
the number of degrees of freedom for the fit), 
for an ``$r^{1/4}$'' De Vaucouleurs law 
($I = I_e\times {\rm exp}(-7.67 [(r/r_e)^{1/4} - 1])$)
with parameters $r_e = 1\farcs5 \pm 0\farcs7$ and 
$I_e = 24.9 \pm 0.4$ mag/arcsec$^2$. Figure 6
is a comparison of our profile data with our DeVaucouleurs profile. 
We found a better fit ($\chi^2/DOF \sim 0.8$) 
for a Hubble-law profile ($I = I_0/(1+(r/r_0)^2)$), and a worse fit 
($\chi^2/DOF \sim 2.7$) for an isothermal profile 
($I = I_0/(1+(r/r_0)^2)^{1/2}$). These results are similar to those
obtained by van Dokkum \& Franx (1996) for E galaxies 
observed with HST in the cluster of galaxies 
${\rm CL}\,0024+016$, at $z = 0.4$. 
Such a similarity is a clue that G may be an elliptical galaxy at 
$z\sim 0.4$. 

\section{Gravitational lens models}

As an application of the astrometry and photometry of \S 3, we 
attempted to estimate parameters of lens models that accounted
for the following observed properties of \mg: 
(1) the relative coordinates $x_i$ and $y_i$  of the images of the core 
and of the center of brightness of G
($i =$ A1, A2, B, C and G); (2) the ratios of core 
image fluxes $F_i/F_j$, with 
$i,j =$ A1, A2, B and C for $i< j$; and 
(3) the shape and location of the arc.

It is likely that stars in G do not affect the radio flux ratios 
by microlensing (Witt, Mao \& Schechter 1995, hereafter 
WMS95). As in other lensed systems (e.g., 2237+0305, see
Falco et al. 1996, and references therein), the optical flux ratios are 
probably affected by microlensing. Indeed, the optical and radio flux ratios
are quite different (e.g., KH93): 
for example, the ratio A1/A2 is $\sim 2.0$ in $I$ 
(see Table 3), and $\sim 1.1$ at 8 GHz. 
We attempted to account for the extended emission of the arc
pixel-by-pixel, but we arrived at the conclusion that it is 
too faint to contribute useful constraints based on our data 
(compare with Ellithorpe et al. 1996). 
Thus, we initially used only 
the coordinates of the images of the core and of G in Table 2 
and we excluded the flux ratios as constraints for model fitting. 

We also considered object X, which might be: (1) in the foreground
of and far from G, and therefore unrelated to the lens system; (2) at about
the same redshift as G, implying that its contribution to the lensing
must be estimated (note that the $R-I$ color of X lies between 
that of G and that of the images of the core, and therefore 
is marginal evidence against X sharing the redshift of G or of the QSO);
and (3) in the background of G, 
and singly-lensed. We consider the last two possibilities below. 

We assumed for most of our models that G is the sole contributor to
the lensing in the \mg\ system (WMS95),
save for including object X in one case (see below). 
We assumed that the center of brightness of G is also its center of mass.
We used an iterative fitting algorithm (with the
software suite ``lensmod'', see Leh\'ar et al. 1993, modified 
to operate in the image plane) that 
treats lensed images as points instead of extended objects. 
The $\chi^2$ goodness-of-fit measure for our models is a sum of squares 
of position differences (each weighted by 
the squared inverse of its standard error) for 
the images of the core. We also added 
a similarly-weighted contribution to $\chi^2$ from the coordinates of G. 
We found the ``best-fit'' lens model 
by direct search for minima of $\chi^2$ in parameter space, with
the ``amoeba'' algorithm (see, e.g., Press et al. 1992). 
We analyzed in this way a series of different models (A $-$ D), 
each based on a different, but simple potential. 

In our first model, A, we assumed a constant M/L$_I$ ratio for G
(see Schneider et al. 1988), and used 
the elliptical isophotes for G (\S 3). We adjusted 
the M/L$_I$ ratio, represented by the radius of the Einstein ring
of G, as well as $x_G$, $y_G$ 
and the mean position angle $\Psi$ of the light distribution of G. 
Thus, for model A we have 4 parameters and 
8 constraints. 
Figure 7 shows the caustic and critical
curves, and image and source positions for model A. 
The fit is poor, as indicated by the final $\chi^2/DOF$ of $\sim 13$.

For the other 
models, we assumed an elliptically-symmetric, two-dimensional  
potential $\phi$ for G, as in Blandford \& Kochanek (1987, hereafter BK87). 
The adjustable parameters were $x_G$ and $y_G$; 
the angular core radius $\theta_c$;  
the radius, $\theta_E$, of the Einstein ring for 
G; the ellipticity $e = 1 - b/a$, where $b/a$ is the 
axis ratio of the mass distribution, and the position angle $\Psi$
of the major axis of the mass distribution (on the sky, E of N). 
We did not adjust 
the power-law exponent $P=2-2\alpha$ of $\phi$, where $\alpha$ is the
``hardness" of BK87 (fitting experiments showed that we cannot 
constrain the value of $P$ usefully). 

For model B, we fixed $P=1$ and $\theta_c=0$, which corresponds to a 
singular isothermal profile; thus, we had 5 parameters and 8 constraints. 
Figure 7 shows the caustics and critical
curves, as well as the image and source positions for model B. 
Compared to model A, model B yielded an improved 
$\chi^2/DOF$, $\sim 8$. 

Model C was the same as model B, except that we set $e=0$, and added 
an external shear parameterized by $\gamma_e$ as in
WMS95, measured in units of the critical surface mass density $\Sigma_c$. 
Model D was the same as model C, except that we set 
$P=2$, which corresponds to a 
point mass. With both of these models, we obtained improved 
fits, with final $\chi^2/DOF$ of $\sim 5.4$ and $\sim 4.5$, respectively
(see Figure 7).
Table 4 shows the parameter values 
for models $B-D$. The errors quoted in the table are the changes
in the parameters that separately yield an increase of $\chi^2/DOF$ of 1. 
These errors do not account for the systematic effects that 
cause $\Delta\chi^2 > 1$.

%*********************************

The 15 GHz flux ratios of the images of the core were recently shown to be 
consistent with no fluctuations, at the $\sim 3$\% rms level 
over a period of 6 months 
(Moore \& Hewitt 1996, hereafter MH96). 
We re-estimated the parameter values 
in each of our models A-D, after including the mean 
KH93 flux ratios as constraints, with $\sim 3$\% errors. 
We found that $\chi^2/DOF$ for the model A$-$D 
fits worsened by 5, 12, 40 and 40\%, respectively. 
These new parameter values led to 
predicted image flux ratios that contributed insignificantly to
$\chi^2$, whereas the core image coordinates contributed
$2-3$ times more to $\chi^2$ than for the original parameter
values in models A-D. 
Thus, we are unable to account in detail for the full set 
of measured properties of the images of the core. 

Although we did not include them as constraints to
our fits, we are able to account qualitatively for the arcs. 
Figure 8 shows the image and source planes for 
model C, with simulated source components: the four QSO images
are produced by a single compact source inside the astroid caustic, and 
the arc is produced by an extended source that lies just inside
the caustic. 
In VLBI observations (PP96), extended
radio emission (roughly aligned with the
arc emanating from B) is apparent near image B, with no 
counterpart near images A1
and A2. Such a configuration is consistent 
with the source of extended radio emission being just outside the caustic
(see Figure 8). 

The arc and the images of the core share geometric properties 
(the separation of the arc and core sources 
is $\sim 200 h^{-1}$ pc, according to our model C), 
but their measured colors are dissimilar. 
Their mean radial distance from the center of G
is $1\farcs2 \pm 0\farcs2$, and the arc 
extends over a comparable distance, $\sim 1\farcs5$. 
The colors of the images of the core 
have a mean of $R-I = 2.2 \pm 0.1$ mag, and differ from 
the mean $R-I = 1.3 \pm 0.3$ mag color for the arc. 
Thus, if the arc and the core sources 
had the same color, the measured 
color differences would argue in favor of reddening in
the host galaxy of the QSO, instead of in the lens galaxy. 
However, it is possible, for example,  that 
the arc is a distorted view of an HII region
with no physical association with the QSO core, and 
with a bluer color. If the lens galaxy G were 
at $z = 0.5$, the ``most likely'' value estimated by Kochanek 
(1992), then extinction in G 
could account qualitatively for the optical color differences 
among A1, A2, B and C, as well as for the IR magnitudes and for the 
unusual line ratios found in this system (LEJT95). The color differences 
would arise from extinction in G that is sufficiently 
patchy to yield such differences 
on a length scale of $\sim 0\farcs4$, the smallest 
separation between the images of the core. 
We find no variation in the color of the arc as a function of position
along it, but only at a modest $\sim 0.3$ mag level of significance, 
larger than the color differences among A1, A2, B, and C (Table 2). 
If a lack of variation were confirmed, say, at the $\sim 0.03$ mag level, 
it would become a much 
stronger argument against patchy extinction in G, and in favor
of extinction in the host galaxy of the QSO, rather than in G. Further
high-resolution IR studies of the system will provide 
improved constraints on the location of the extinction. With 
B. McLeod, we intend
to continue optical and IR studies both from the ground and with HST.

Because it appears aligned with the continuation of the arc past B,
object X seems morphologically associated with \mg. If it were not in
the foreground of G (the least interesting possibility), we 
could have: (1) 
$z_{\rm X} \sim z_{\rm QSO}$, (2) $z_{\rm G} < z_{\rm X} < z_{\rm QSO}$,
or (3) $z_{\rm X} \sim z_{\rm G}$.
In the first case, X could be expected to have a counter-companion
(a counter-arc in our example, or at least a second image 
likely to be at least as bright as X) 
that we did not detect, and that could not have been
missed by our observations (see Figure 8, dotted contour).
In the second case, the redshift of X is not likely to exceed that of 
G significantly, because the Einstein ring radius $\theta_E({\rm X})$
for X depends strongly on $z_{\rm X}$. We can illustrate this property 
by assuming that the mass distribution of G is 
a singular isothermal sphere, an approximation that preserves
the characteristic angular scale of the lens system. 
For such a distribution, the radius of the multiple-image region 
for a source at $z_{\rm X}$ is $2 \,\theta_E({\rm X})$.
Within this approximation, for X to be a single image, 
$\theta_E({\rm X}) < \theta({\rm X})/2$, 
where $\theta({\rm X}) \sim 1\farcs5$ 
is the angular distance from X to 
the center of brightness of G (Tables 1 and 2). However, from Table 4, 
$\theta_E \sim 1\farcs2 \approx \theta({\rm X})$; 
which is approximately equivalent to $\theta_E({\rm X}) \ltorder \theta_E/2$. 
Thus, if $z_{\rm G} \approx 1$, then 
$z_{\rm X}-z_{\rm G} \approx0.4$, with this difference 
being smaller for any other $z_{\rm G}$.
In the third case, we assumed that 
the M/L$_I$ ratio of X is equal to that of G, and fitted a modified model A
that included X in this fashion. We found 
$\chi^2/DOF\sim 41$ for this model, a very significantly worse fit
than for our other models. 
Because X is clearly not point-like, we infer that it may 
be a dwarf galaxy near the redshift of G, or in its foreground.

\section{Conclusions}

We conclude first that lensing is compellingly confirmed
for \mg, in spite of the lack of a 
measured redshift for the lens galaxy.
It seems very probable that 
the arc that we discovered is a highly magnified and distorted image of 
a patch of the source near its core. Together with 
the morphology of the images of the core, as refined by 
our observations, the arc is strong evidence in favor of lensing.

Our observations yielded positions for all 4 known
images of the QSO central core,
with standard errors of $0\farcs02$, and for the lens galaxy G with
standard errors of $0\farcs05$; 
our $R$ and $I$ photometry for these objects has
standard errors $\sim 0.01$ mag. The photometry of the arc is much less
accurate ($\sim 0.3$ mag standard errors), because it is extremely faint, especially
in the $R$ filter. 

The brightness distribution of G is well represented by a De
Vaucouleurs radial profile with effective radius 
$r_e = 1\farcs5 \pm 0\farcs7$ and surface brightnesss 
$I_e = 24.9 \pm 0.4$ mag/arcsec$^2$, 
and by elliptical isophotes with ellipticity 
$e_G = 0.20 \pm 0.02$ and position angle $\Phi_G = 71^\circ \pm 5^\circ$.
Thus, we conclude that G is most likely an elliptical galaxy, whose redshift
unfortunately remains unknown.

Our lens models qualitatively account for both the compact QSO images
and the arc. However, we could not find any models that were 
simultaneously consistent in detail with the HST positions 
of the images of the core 
and with their (unvarying, and unaffected by microlensing) 
radio flux ratios, as given by KH93. Monitoring of the images of \mg\ 
at radio frequencies shows variations at 15 GHz of only $\leq 3.5$\% 
on time scales of 6 months (MH96). Thus, the discrepancy 
between observed and predicted flux ratios exhibits 
a serious deficiency of our models.

There are significant differences between the color of the 
images of the core QSO source, $R_{\rm core}-I_{\rm core} = 2.2 \pm 0.1$ mag, 
and that of the arc, $R_{\rm arc}-I_{\rm arc} = 1.3 \pm 0.3$ mag.
The latter color appears to be invariant along the length of the arc 
between images A2 and B, but at a modest level of significance.
If such invariance were confirmed with high significance, we 
might be able to usefully test detailed extinction models.  
Reddening in G would likely be patchy at the same level as 
the observed variations among the images of the core 
(RMS $\sim 0.13$ mag) over the length of the arc, and should affect
the latter, as well as the former. Therefore, 
if our measurements were confirmed with greater accuracy, they would 
suggest that (at least) the bulk of the 
reddening occurs in the host galaxy of the QSO, rather than in G. 
The large discrepancies between the $R$ ($2.50 \pm 0.04$), $I$ 
($2.11 \pm 0.01$) and radio ($1.13 \pm 0.03$) flux ratios A1/A2 would then 
imply that microlensing is important on at least one
of the paths from the core to images A1 and A2, as suggested in WMS95. 
Optical/IR monitoring observations could reveal microlensing, but it is
possible that A2, for instance, is 
and would remain in a ``low'' state for a period
of years; disentangling the possibilities 
could then require a very long temporal baseline. 

In each of 
our models B-D, the estimated ellipticity of the mass distribution (or
the amount of shear) is large, about twice that of the light from G. 
The ellipticities of the 
critical curves straddled by the images of the core 
in Figure 7 are indicative
of the discordant ellipticities of the light and matter in G. 
The position angles of our mass models are misaligned 
by only $\sim 6^\circ$ (within our standard errors)
with the orientation of the light distribution of G. Such 
ellipticity discrepancies are common for models of quadruple lens systems
(Kochanek 1996a). Possible explanations are external shears
due to, e.g., nearby objects, or local effects, such as 
triaxiality (Kochanek 1996b, private communication). 
``Reasonable'' modifications of the profile of the mass distribution of G 
are not expected to eliminate such discrepancies. 

Determination of the redshift of the lens galaxy G is a very 
important missing ingredient for further analysis of \mg. 
Such an estimate could be used to derive useful physical properties
of the lens and of the lensed QSO, such as masses and
linear sizes. For example, for 
models C and D, given the errors in the 
estimates of the positions of the core images,
the  mass of G within a radius
of $\sim 1\farcs2$ is $M_G = 2.1 \pm 0.1 \times 10^{11} M_\odot$,
for $z_G=0.5$. Also, 
with the redshift of G measured and combined 
with estimates of other model parameters, one could infer useful limits
for the linear sizes of the core and the arc source (WMS95), and 
for their separation on the sky. 

The nature of object X is not yet determined. Its extended 
structure suggests that it is a galaxy. Because it is singly imaged,
X is unlikely to 
be at the redshift of the QSO. If it were at the lens redshift
or closer, it could not be accounted for with lens models that 
assume the same M/L$_I$ ratio as for model A. However, were X at a redshift
intermediate between the lens and the QSO, it could be a major deflector
component. Difficult even for Keck-class telescopes, 
but very useful, would be a measurement of the redshift of X.

We gratefully acknowledge the support of ST grant G5505 and NSF grant
AST93-03527. We thank C. Kochanek, B. McLeod, P. Schechter and
J. Huchra for useful comments and discussions. We also thank the
anonymous referee for careful, helpful suggestions.

\centerline{FIGURE CAPTIONS}

\noindent Figure 1: 
PC1 5\farcs8-field centered on \mg; obtained with the F675W filter ($R$).
To account for photon
statistics, we used gain $g = 7.12 {\rm e}^-$/count 
and read-out noise $r = 5.24 {\rm e}^-$ for our individual exposures.
We converted from pixel to equatorial coordinates 
$\alpha$ and $\delta$ for equinox J2000 with 
the IRAF STSDAS package. The software corrects for geometric distortions of
WFPC2, and uses the appropriate transformations to sky
coordinates from parameters incorporated in the image headers. 
The orientation and scale of the field on the sky are indicated by the arrows.
The images 
of the core, object X and the lens galaxy G are all labeled.\\

\noindent Figure 2:  PC1 23\farcs3-field surrounding \mg; 
obtained with the F675W filter. The orientation and scale 
of the field on the sky are indicated by the arrows.
A satellite trail is still visible, having crossed one of our exposures
from W to E.\\

\noindent Figure 3: 
PC1 5\farcs8-field centered on \mg; obtained with the F814W filter ($I$).
The orientation and scale of the field on the 
sky are as in Figure 1. The emission connecting A2 and B appears
fainter than in Figure 1, 
because the arc is bluer than the images of the core.\\

\noindent Figure 4: PC1 23\farcs3-field surrounding \mg; 
obtained with the F814W filter. 
The orientation and scale of the field on the sky are as in 
Figure 2.\\

\noindent Figure 5:
Contour plot of the MEM-deconvolved $R$ image. Note that this plot is rotated
relative to the orientation of Figure 1; it is aligned with the equatorial
coordinate axes.\\

\noindent Figure 6:
Circles with error bars are the derived radial profile of G, as a function
of $r^{1/4}$, where $r$ is the semi-major axis in arcsec. The long (short)  
dashes 
are the ``best-fit'' De Vaucouleurs profile to the measured profile, 
before (after) this profile is convolved with the PSF radial profile.\\

\noindent Figure 7:
Caustics (solid) and critical curves (dots) for models A$-$D.
The positions of the images of the core on the image
plane are indicated with $+$ signs. 
Solid ellipses are the core image position errors,
traced to the source plane for each model. The
goodness-of-fit of each model is indicated by the closeness 
to each other of these ellipses; they 
would all coincide in position for a perfect fit. Their sizes 
in each model 
are inversely proportional to the magnification of each core image.\\

\noindent Figure 8:
Caustic (top) and critical curve (bottom) for model C. 
Objects labeled Q, A, J and X on the source plane (top) 
correspond to similarly labeled objects on the image 
plane (bottom). Source Q and its images (dark)
are unresolved, and correspond to the QSO core and
its images. Sources A (shaded)
and J (solid circles and ellipses) are resolved, and correspond to the 
optical arc (our observations) and the radio jets (PP96), respectively.
Source X (dashed, empty circle) would be multiply imaged as
object X and the elongated dashed arc (bottom), if it were at $z\sim 2.6$.\\

\clearpage

\centerline{TABLE CAPTIONS}

\noindent Table 1: Right ascension and declination offsets
from C, and $R$ magnitudes of the components of \mg, including
object X, 3 stars, and a possible group of galaxies. 
The coordinates of the
group of galaxies are the means of the 
coordinates of the 3 galaxies visible in our
exposures. The standard error for each coordinate is $\sigma_{\rm x,y}$.\\

\noindent Table 2: As Table 1, for $I$ band properties, 
with the addition of $R-I$ colors. \\

\noindent Table 3: Flux ratios of the images of the core
in $R$ and $I$. The combination A1$+$A2 was used to allow comparison
with the ground-based estimates of SM93 and AVHM94 (see also
LEJT95, Table 1). \\

\noindent Table 4: Model parameter values, standard
errors, and $\chi^2/DOF$ for \mg.\\

\clearpage

\begin{table}
\begin{center}
\begin{tabular}{lr@{.}lr@{.}lr@{.}lr@{.}lr@{.}l}
\multicolumn{11}{c}{Table 1}\\
\tableline
Object  &
\multicolumn{2}{c}{$\Delta\alpha_{\rm J2000}$} &
\multicolumn{2}{c}{$\Delta\delta_{\rm J2000}$} & 
\multicolumn{2}{c}{$\sigma_{\rm x,y}$} &
\multicolumn{2}{c}{$R$}&
\multicolumn{2}{c}{$\sigma_R$} \\
        &    
\multicolumn{2}{c}{(arcsec)}     & 
\multicolumn{2}{c}{(arcsec)} & 
\multicolumn{2}{c}{(arcsec)} & 
\multicolumn{2}{c}{(mag)} & 
\multicolumn{2}{c}{(mag)} \\
\tableline
A1    &  1&92  &   -0&30  & 0&02 & 22&760  &    0&008 \\
A2    &  2&05  &    0&09  & 0&02 & 23&756  &    0&016 \\
B     &  1&32  &    1&63  & 0&02 & 23&488  &    0&012 \\
C     &  0&     &    0&   & 0&02 & 24&258  &    0&022 \\
G     &  0&90  &    0&32  & 0&05 & 23&227  &    0&046 \\
X     &  0&46  &    1&84  & 0&05 &$>26$&268 &    0&063 \\
Star 1&  5&89  &   10&77  & 0&02 &  21&887  &    0&004 \\
Star 2& -0&33  &   -16&92  & 0&02 &  22&997  &    0&009 \\
Star 3& 17&59  &  -13&40  & 0&02 &  23&97  &    0&02 \\
Group & -3&30  &  -2&87  &  0&08 & 24&14  &    0&11 \\
\tableline
\end{tabular}
\end{center}
\end{table}

\clearpage

\begin{table}
\begin{center}
\begin{tabular}{lr@{.}lr@{.}lr@{.}lr@{.}lr@{.}lr@{.}l}
\multicolumn{13}{c}{Table 2} \\
\tableline
Object  &
\multicolumn{2}{c}{$\Delta\alpha_{\rm J2000}$} &
\multicolumn{2}{c}{$\Delta\delta_{\rm J2000}$} & 
\multicolumn{2}{c}{$\sigma_{\rm x,y}$} &
\multicolumn{2}{c}{$I$}&
\multicolumn{2}{c}{$\sigma_I$} & 
\multicolumn{2}{c}{$R-I$}\\
        &    
\multicolumn{2}{c}{(arcsec)}     & 
\multicolumn{2}{c}{(arcsec)} & 
\multicolumn{2}{c}{(arcsec)} & 
\multicolumn{2}{c}{(mag)}& 
\multicolumn{2}{c}{(mag)} \\
\tableline
A1    &  1&92  &   -0&29  & 0&02 &  20&595  &    0&003 & 2&17\\
A2    &  2&05  &    0&10  & 0&02 &  21&407  &    0&004 & 2&35\\
B     &  1&32  &    1&64  & 0&02 &  21&363  &    0&004 & 2&13\\
C     &  0&     &    0&   & 0&02 &  22&212  &    0&007 & 2&05\\
G     &  0&86  &    0&36  & 0&05 &  21&296  &    0&014 & 1&93\\
X     &  0&46  &    1&84  & 0&05 &  24&769  &    0&063 &$>1$&50\\
Star 1&  5&88  &   10&78  & 0&02 &  21&271  &    0&004 & 0&62\\
Star 2& -0&33  &  -16&91  & 0&02 &  22&195  &    0&007 & 0&80\\
Star 3& 17&58  &  -13&39  & 0&02 &  22&43  &    0&008 & 1&54\\
Group &  -3&27  &  -2&67  & 0&06 &  23&16  &    0&09 & 0&98\\
\tableline
\end{tabular}
\end{center}
\end{table}

\clearpage

\begin{table}
\begin{center}
\begin{tabular}{cr@{.}lr@{.}l}
\multicolumn{5}{c}{Table 3}\\
Ratio  &
\multicolumn{2}{c}{$R$} &
\multicolumn{2}{c}{$I$}  \\
\tableline
A1/A2    &  2&50 $\pm$ 0.04  &   2&11 $\pm$ 0.01\\
(A1$+$A2)/B &  2&72 $\pm$ 0.06 &   2&98 $\pm$ 0.07 \\
C/B     &  0&49 $\pm$ 0.01 &    0&46  $\pm$ 0.01\\
\tableline
\end{tabular}
\end{center}
\end{table}

\clearpage

\begin{table}[h]
\mbox{\small
\begin{minipage}[h]{6.5in}
\begin{tabular}{lr@{.}lr@{.}lr@{.}lr@{.}lr@{.}lr@{.}lc}
\multicolumn{14}{c}{Table 4}\\
\tableline
 &
\multicolumn{2}{c}{$x_G$} &  
\multicolumn{2}{c}{$y_G$} & 
\multicolumn{2}{c}{$\theta_E$}&
\multicolumn{2}{c}{$\gamma_e$} & 
\multicolumn{2}{c}{$e$} & 
\multicolumn{2}{c}{$\Psi$}&
$\chi^2/DOF$\\
Model &
\multicolumn{2}{c}{(arcsec)} &  
\multicolumn{2}{c}{(arcsec)} & 
\multicolumn{2}{c}{(arcsec)}&
\multicolumn{2}{c}{} & 
\multicolumn{2}{c}{} & 
\multicolumn{2}{c}{($^\circ$ E of N)}&\\
\tableline
B  & 0&81 $\pm$ 0.04 &  0&31 $\pm$ 0.02 &  1&17 $\pm$ 0.01 & 
\multicolumn{2}{c}{} & 0&37 $\pm$ 0.04 & 78&3 $\pm$ 0.5 &8.0\\
C  & 0&83 $\pm$ 0.04 &  0&31 $\pm$ 0.02 &  1&18 $\pm$ 0.01 &   
0&12 $\pm$ 0.01  &\multicolumn{2}{c}{}  & 77&5 $\pm$ 0.5&5.5\\
D  & 0&84 $\pm$ 0.03 &  0&30 $\pm$ 0.02 &  1&16 $\pm$ 0.01 &
 0&25 $\pm$ 0.03  & \multicolumn{2}{c}{} & 77&4 $\pm$ 0.5&4.5\\
\tableline
\end{tabular}
\end{minipage}}
\end{table}

\clearpage
\begin{figure}
\plotone{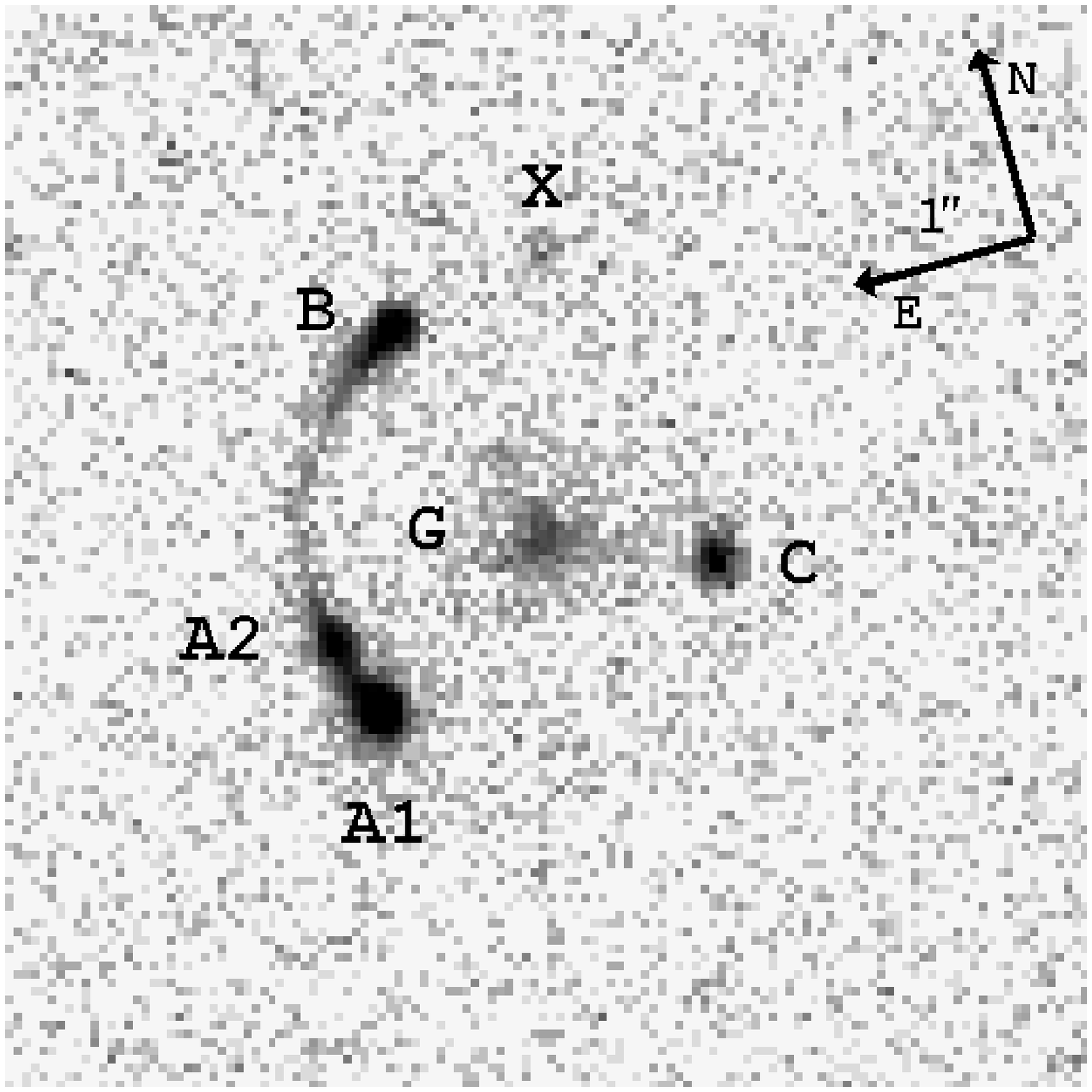}
\end{figure}

\clearpage

\begin{figure}
\plotone{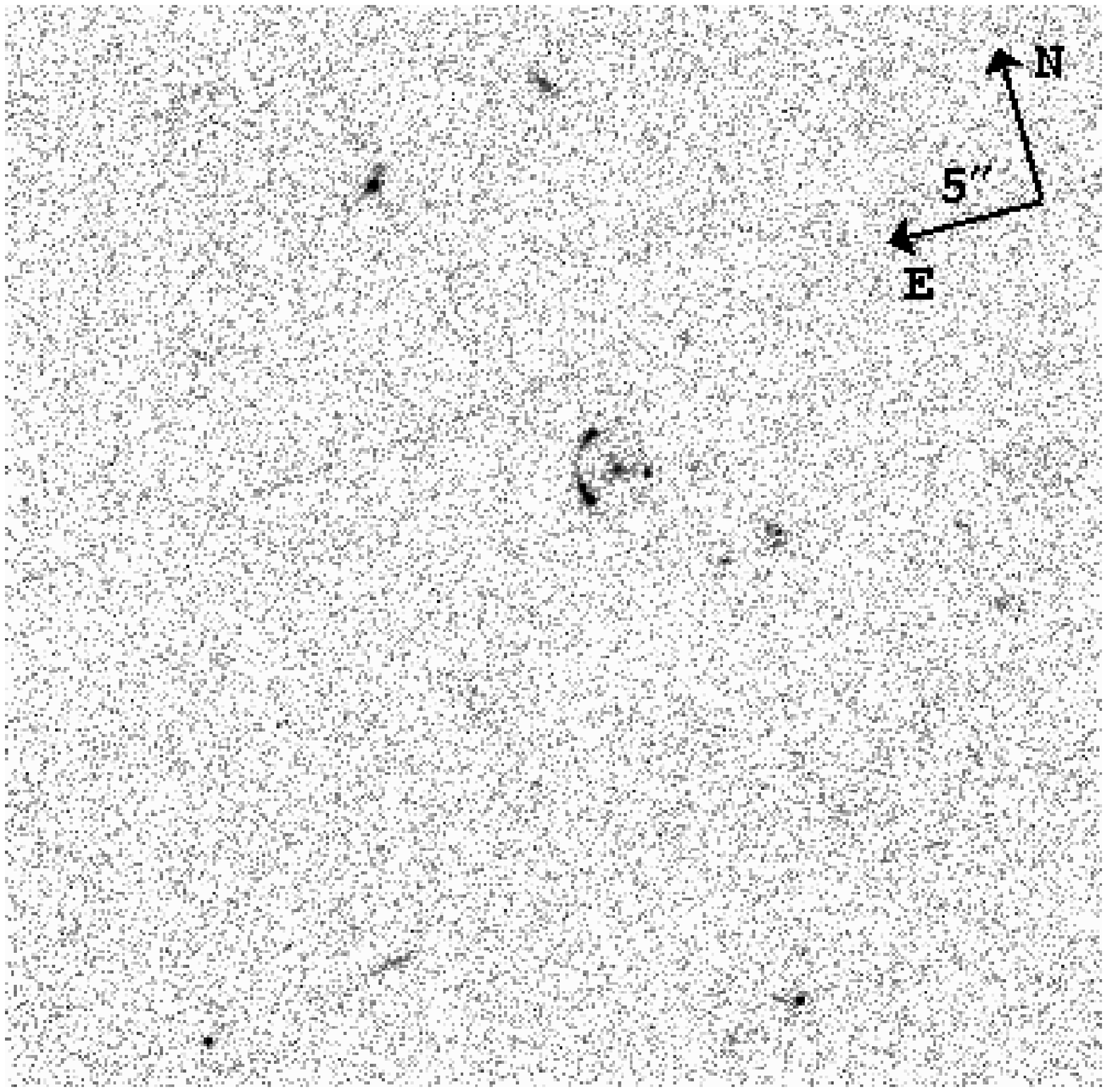}
\end{figure}

\clearpage

\begin{figure}
\plotone{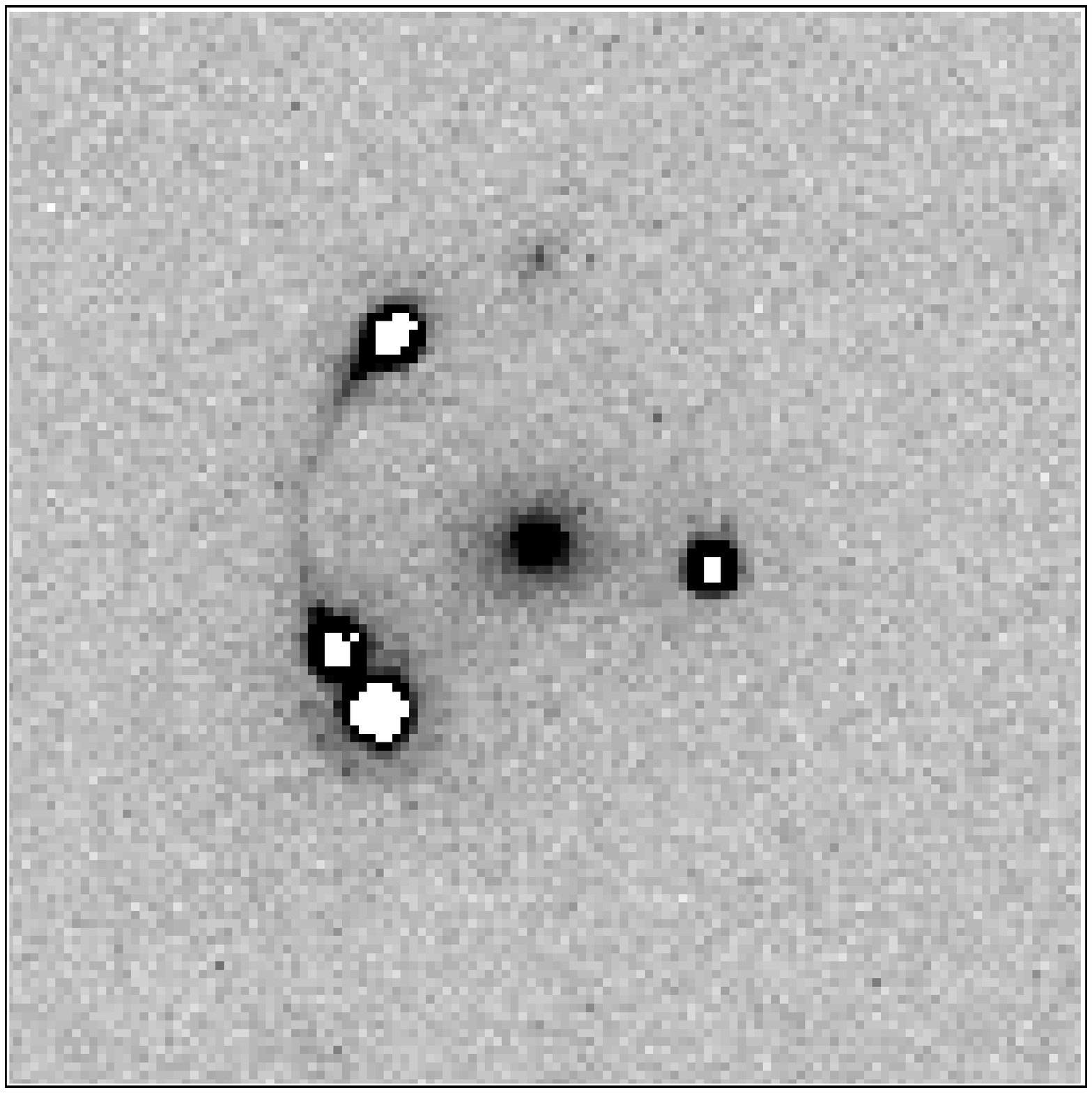}
\end{figure}

\clearpage

\begin{figure}
\plotone{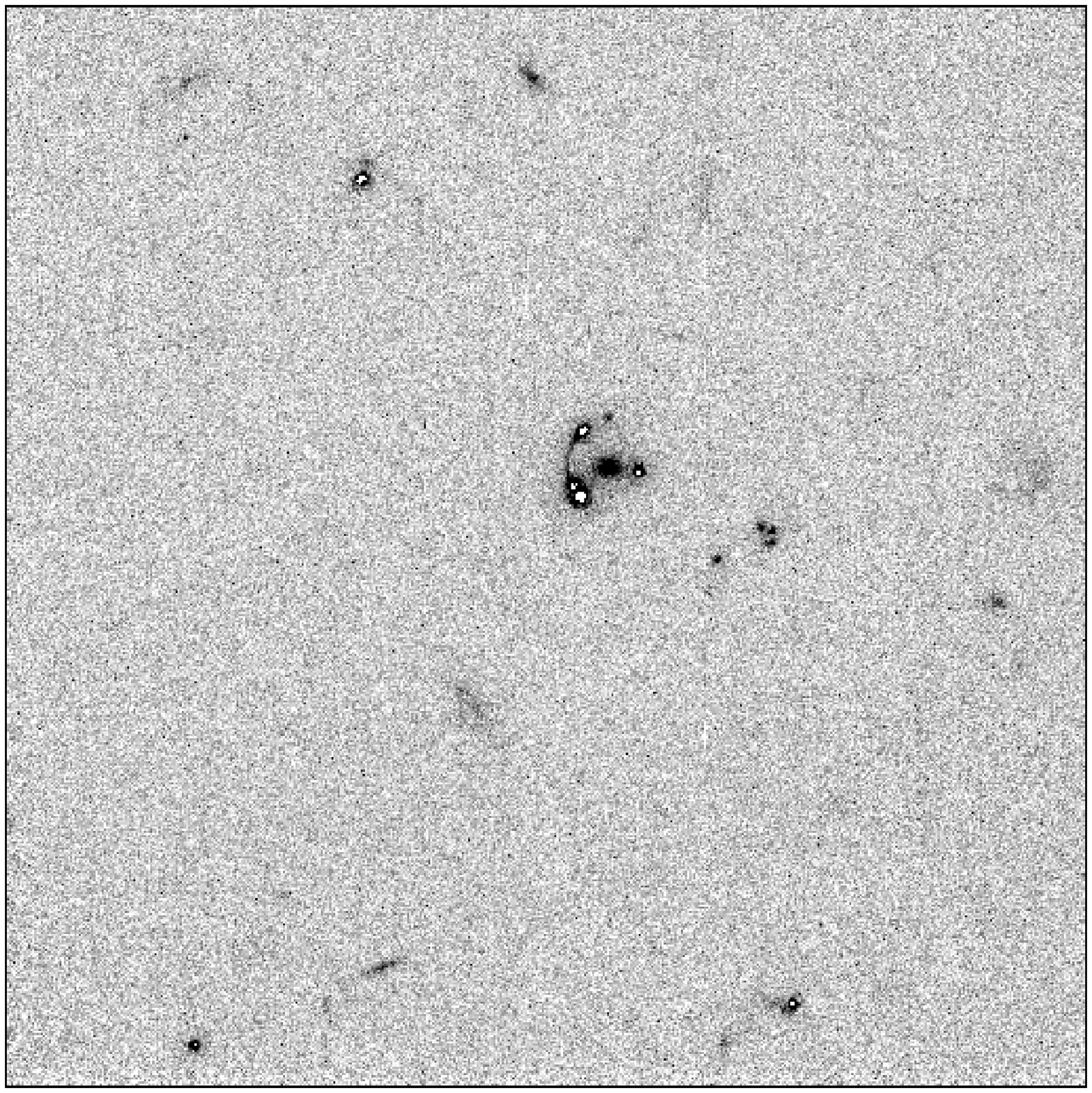}
\end{figure}

\clearpage

\begin{figure}
\plotone{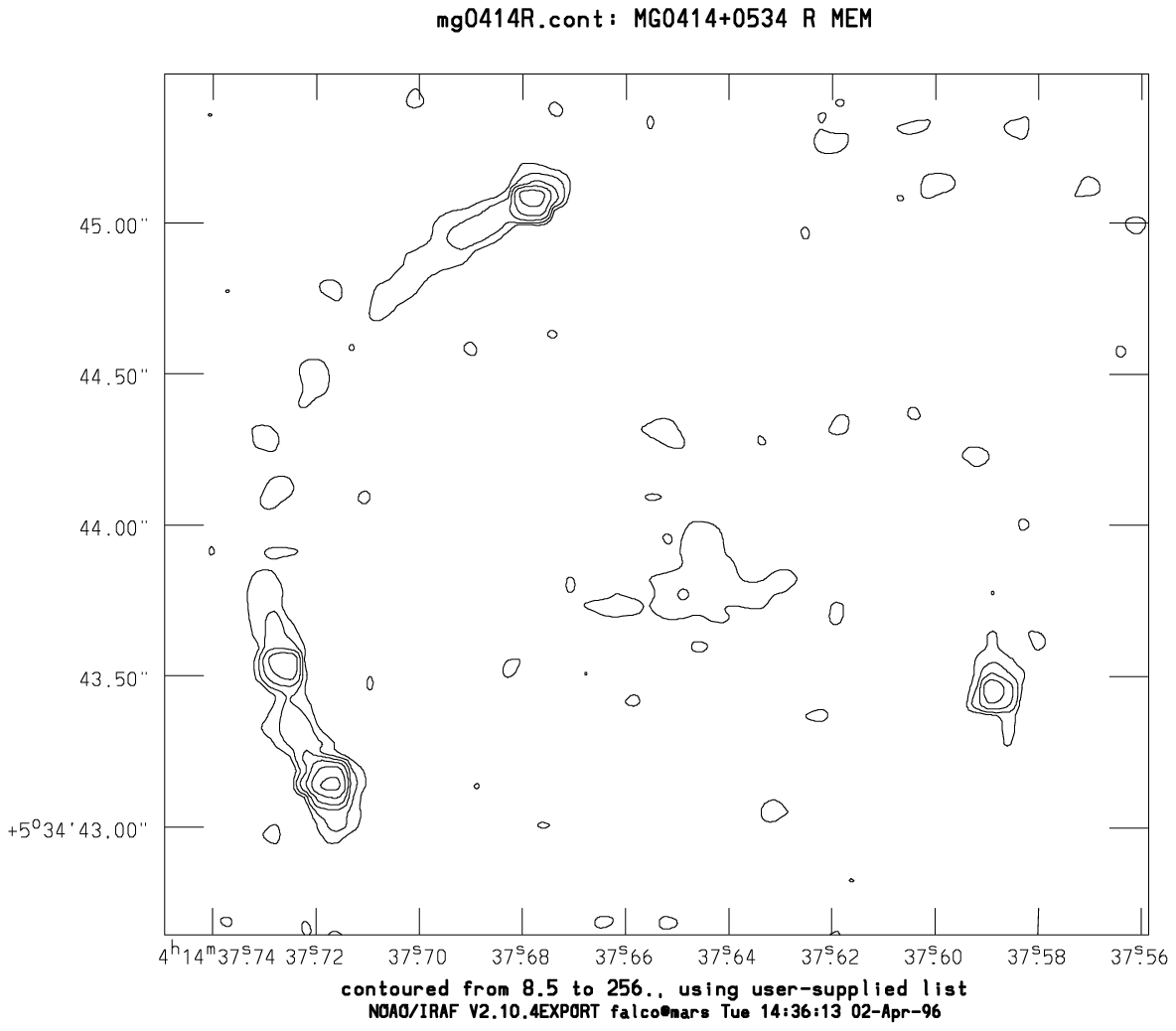}
\end{figure}

\clearpage

\begin{figure}
\epsscale{0.75}
\plotone{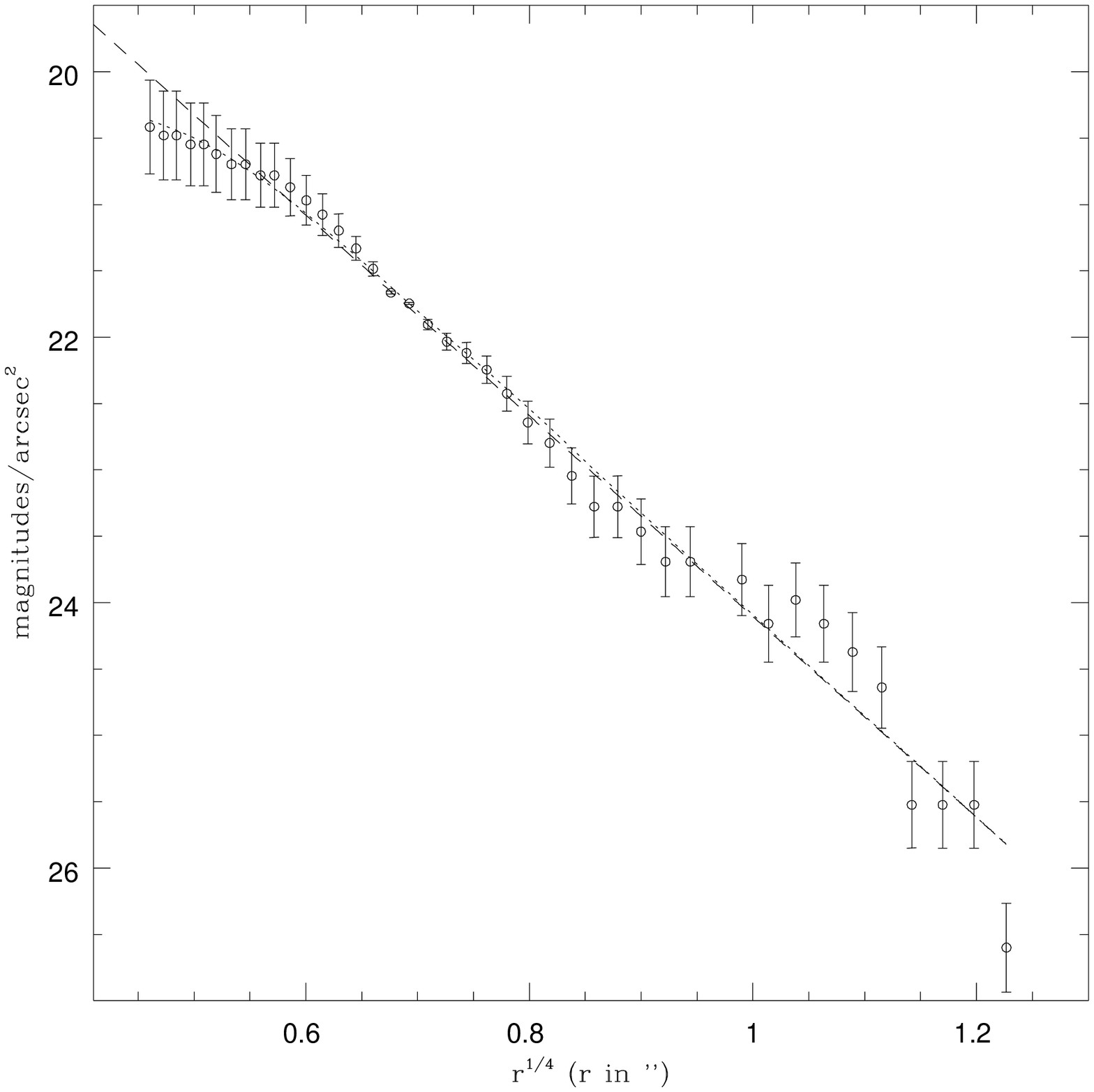}
\end{figure}

\clearpage

\begin{figure}
\epsscale{1.0}
\plotone{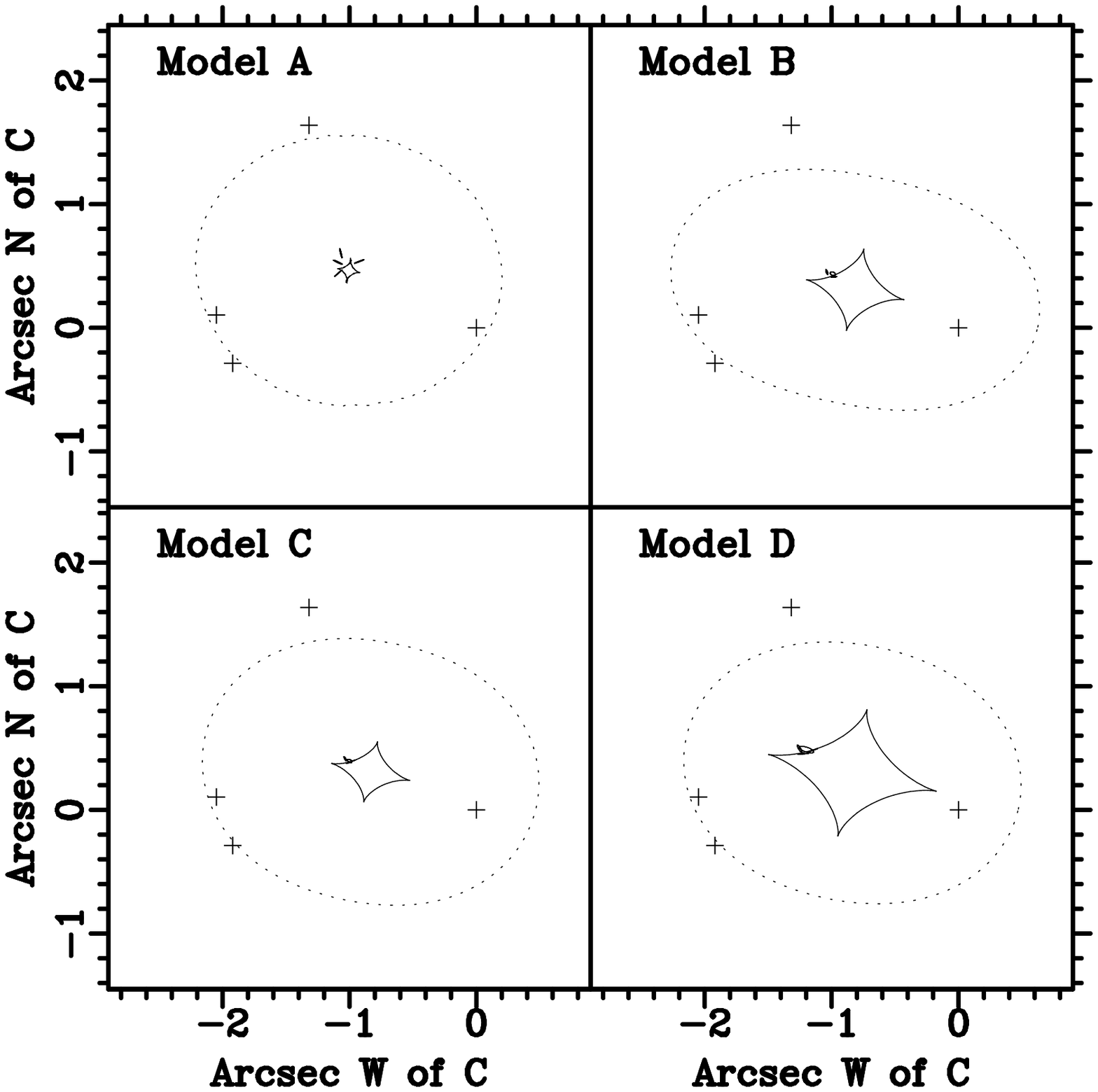}
\end{figure}

\begin{figure}
\epsscale{1.0}
\plotone{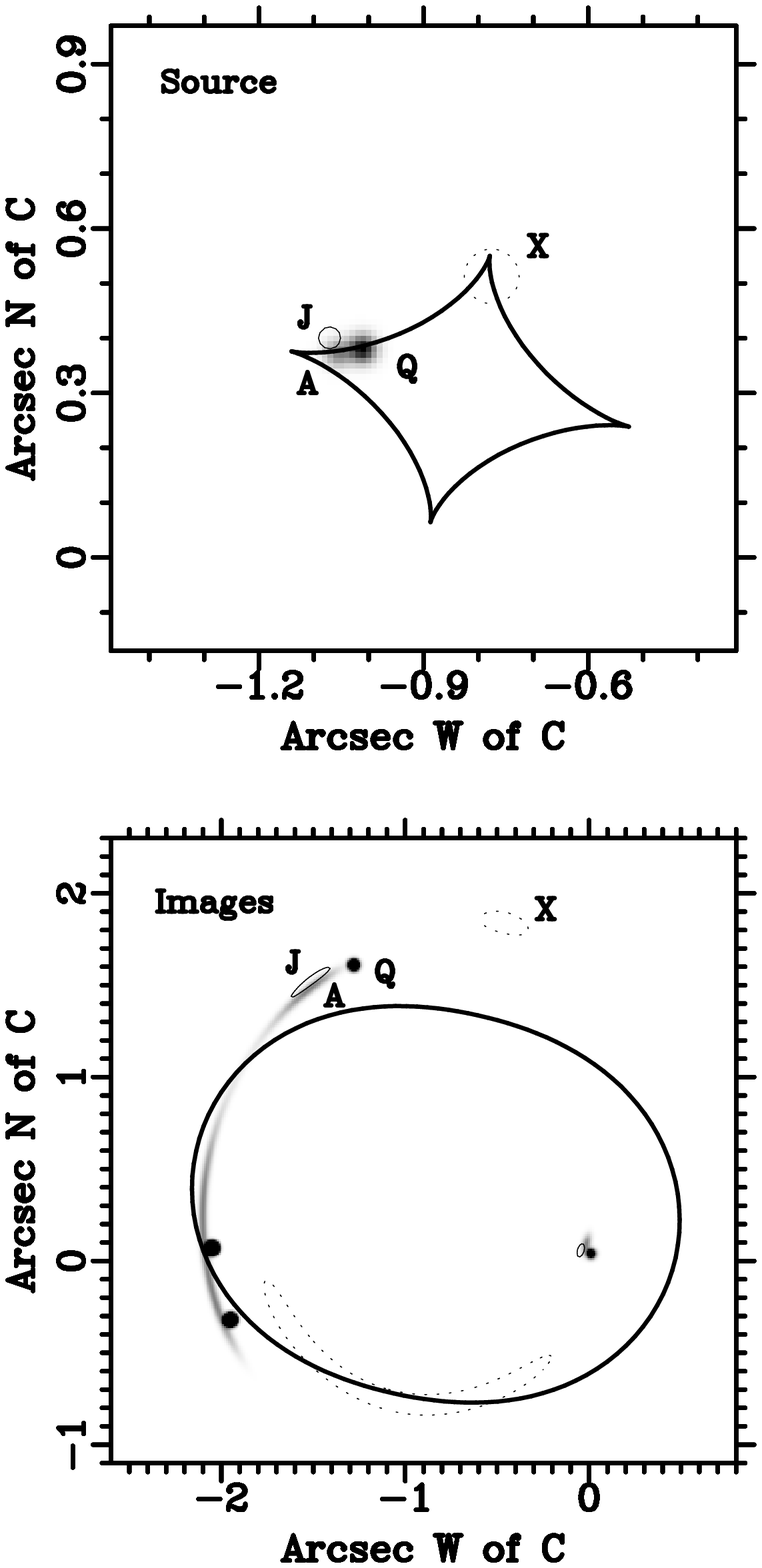}
\end{figure}

\end{document}